\documentclass[conference,a4paper]{IEEEtran}

\usepackage[utf8]{inputenc}
\usepackage[T1]{fontenc}
\usepackage{url}
\usepackage{ifthen}
\usepackage{cite}
\usepackage[cmex10]{amsmath} % Use the [cmex10] option to ensure complicance
\usepackage{amssymb}
\usepackage{nicefrac}
\usepackage{cancel}
\usepackage{enumerate}
\usepackage{theorem}
\theoremstyle{plain}
\newtheorem{theorem}{Theorem}
\newtheorem{example}{Example}
\newtheorem{proposition}{Proposition}

\usepackage{mathabx}
\usepackage{arydshln}
\usepackage{faktor}
\usepackage{stfloats}
\usepackage[dvipsnames]{xcolor}
\usepackage{tikz}
\usetikzlibrary{matrix, decorations.pathreplacing}

\renewcommand{\vec}[1]{\ensuremath{\boldsymbol{#1}}}
\newcommand{\mat}[1]{\ensuremath{\boldsymbol{#1}}}
\newcommand{\N}{\ensuremath{\mathbb{N}}}
\newcommand{\F}{\ensuremath{\mathbb{F}}}

\begin{document}

\title{A Linear Algebraic Approach to Subfield Subcodes of GRS Codes}

\author{%
  \IEEEauthorblockN{Christian Senger and Rohit Bohara}
  \IEEEauthorblockA{Institute of Telecommunications\\
                    University of Stuttgart\\
                    70199 Stuttgart, Germany\\
                    Email: \{senger, bohara\}@inue.uni-stuttgart.de}
}

\maketitle

\begin{abstract}
  The problem of finding subfield subcodes of generalized Reed--Solomon (GRS) codes (i.e., alternant codes)
  is considered. A pure linear algebraic approach is taken in order to derive message constraints that
  generalize the well known conjugacy constraints for cyclic GRS codes and their
  Bose--Chaudhuri--Hocquenghem (BCH) subfield subcodes. It is shown that the presented technique can be used
  for finding nested subfield subcodes with increasing design distance.
\end{abstract}

\section{Introduction}\label{sec:intro}

\emph{Generalized Reed--Solomon (GRS)} codes are among the most well-researched classes of error-correcting
codes. Efficient decoders up to and beyond half their minimum distance are widely available. One shortcoming
of GRS codes is, that their length cannot exceed the size $Q$ of the finite field $\F_Q$ over which they are
defined. It was shown by Delsarte \cite{delsarte:1975} that restricting the codeword components of a GRS
code to a subfield $\F_q\subset\F_Q$ delivers an \emph{alternant} code as introduced by Helgert
\cite{helgert:1972, helgert:1974}. The latter include \emph{Bose--Chaudhuri--Hocquenghem (BCH)}
\cite{bose_ray-chaudhuri:1960}, \emph{Goppa} \cite{berlekamp:1973}, and \emph{Srivastava} codes as special
cases. These codes are defined over the small field $\F_q$ but their length is only restricted by the size
$Q$ of the big field. It is obvious that subfield subcodes can be decoded using decoders for their GRS
parent code.

Our contribution in this paper is a new (up to our knowledge), purely linear algebraic approach for
obtaining subfield subcodes of GRS codes via message constraints that generalize the well known conjugacy
constraints for cyclic GRS codes and their BCH subfield subcodes. Moreover, we show that our approach can be
used in order to find nested subfield subcodes with increasing distance. Such nested codes are important
building blocks of generalized concatenated codes \cite{zyablov_shavgulidze_bossert:1999}.

\section{GRS Codes and their Subfield Subcodes}\label{sec:codes}

Let $\F_Q$ be the finite field with $Q$ elements. For fixed positive integers $n$ and $k$ with $k \leq n
\leq Q$, let $\F_Q[x]^{<k}$ denote the vector space of polynomials in variable $x$ over $\F_Q$ with degree
less than $k$.

For $n$-tuples $\mathcal{A}=(a_0, \ldots, a_{n-1})$  and $\mathcal{B}=(b_0, \ldots, b_{n-1})$ over $\F_Q$,
in which the components of $\mathcal{A}$ are \emph{distinct and nonzero\footnote{The zero locator is usually
allowed in the definition of GRS codes. It is excluded in this paper because of the column multiplier update in
Proposition~\ref{prop:leading_trailing}.}}, and the components of
$\mathcal{B}$ are \emph{nonzero}, the set
\begin{equation}\label{eqn:code}
  \mathcal{C}=\left\{%
    \left(b_0 f(a_0), \ldots, b_{n-1} f(a_{n-1})\right)\;:\; f(x)\in\F_Q[x]^{<k} \right\}
\end{equation}
represents the codewords of a GRS code (called \emph{modified} RS code
in \cite{delsarte:1975}) over $\F_Q$ with locators
$\mathcal{A}$, column multipliers $\mathcal{B}$, length $n$, dimension $k$, and minimum distance $d=n-k+1$,
(MDS property, see, e.g., \cite{macwilliams_sloane:1988}). Note that $f(x) = \sum_{i=0}^{k-1} f_i x^i \in\F_Q[x]^{<k}$ is
the message polynomial, its coefficients can be chosen freely from $\F_Q$.

Several important classes of polynomial evaluation codes emerge by imposing constraints on $\mathcal{A}$,
$\mathcal{B}$, and $f(x)$. Constraints on $f(x)$ are referred to as \emph{message constraints}. For example, if $\F_Q$ contains a primitive $n$th root of unity, i.e., an element
$\alpha$ of multiplicative order $n$, then a \emph{cyclic} GRS code of length $n$ is obtained from locators
$a_i=\alpha^i$ and column multipliers $b_i=\alpha^{i\delta}$ for some integer parameter $\delta$, where
$i=0,1,\ldots,n-1$.  Note that in this case (due to Lagrange's Theorem), $n$ must be a divisor of $Q-1$,
which, in particular, implies that $n$ and $Q$ are coprime. This is a restriction on the possible code
length that depends on the field $\F_Q$.

If $\F_q$ is a proper subfield of $\F_Q$ (so that $Q=q^m$ for some integer $m>1$) then the set
$\mathcal{C}'=\mathcal{C}\cap\F_q^n$ is called a \emph{subfield subcode} of $\mathcal{C}$ over $\F_q$, cf.
\cite{macwilliams_sloane:1988}. It was observed in \cite{delsarte:1975} that subfield subcodes of GRS codes are in fact
\emph{alternant codes} as introduced in \cite{helgert:1972, helgert:1974}.

Note that $\mathcal{C}'$ has the same block length as $\mathcal{C}$, but its
dimension $k'$ is generally smaller than the dimension $k$ of $\mathcal{C}$. It is not generally true that
the design distance $d'$ of $\mathcal{C}'$ is larger than that of $\mathcal{C}$, but certainly $d'\geq d$.
We are of course interested in the cases where $d'>d$.

The subfield subcodes of cyclic GRS codes are the BCH codes. They can be obtained directly from
\eqref{eqn:code} by choosing locators and column multipliers of a cyclic GRS code and additionally making
the coefficients of the message polynomial $f(x)$ satisfy the message constraint given by the conjugacy
constraints
\begin{equation}\label{eqn:conj}
f_{\pi_\delta[i]}=f_i^q,\quad i=0, \ldots, n-1,
\end{equation}
where
\begin{equation*}%\label{eqn:pi}
  \pi_\delta[i] =   qi+(q-1)\delta \bmod n,
\end{equation*}
which holds for canonical encoding as in \eqref{eqn:code} if and only if $\mathcal{C}\subseteq\F_q^n$, cf.\
\cite{blahut:2011, senger_kschischang:2016}. The message constraint restricts the possible choices of the
coefficients of $f(x)$, which is the reason for $k'<k$.

The choice of $\delta$ has a huge influence on the design distance $d'$ of the subfield subcode. For
example the $\F_2$-subfield subcode of the cyclic GRS code over $\F_{64}$ with length $n=63$, dimension
$k=51$, minimum distance $n-k+1=13$, and parameter $\delta=0$ is the BCH code with dimension $k'=10$ and
design distance $d'=13$. Choosing $\delta=24$ instead results in a BCH code with the same dimension but
much larger design distance $d'=27$. The latter is almost optimal considering the currently known upper
bounds on the minimum distance of linear codes (which allow minimum distance at most $28$ for the parameters
at hand). This example is elaborated in \cite{senger_kschischang:2016}.

It is important to note that any decoder that can correct $t[d]$ errors in the original $\F_Q$-code can be
used to decode up to $t[d']$ errors in the subfield subcode. The practically most relevant example of the
error correcting radius $t[\cdot]$ is $t[x]=\left\lfloor\nicefrac{(x-1)}{2}\right\rfloor$.

Our goal in the following two sections is to derive message polynomial constraints similar to
\eqref{eqn:conj} for \emph{arbitrary} GRS codes, not only cyclic ones. In
order to do that we have to leave the polynomial domain and deal with vectors/matrices instead.

\section{Translating Modular Polynomial Multiplication into Vector/Matrix Domain}\label{sec:translate}

It is well known that for $Q=q^m$ with $q\geq 2$ either a prime or a power of a prime and an integer $m\geq
1$ the finite field $\F_Q$ is given (up to isomorphy) by the quotient
\begin{equation*}
  \left(\faktor{\F_q[x]}{p(x)}, +, \odot\right),
\end{equation*}
where the defining polynomial $p(x)\in\F_q[x]$ is irreducible over $\F_q$ and $\deg[p(x)]=m$. W.l.o.g. we
assume $p(x)$ to be monic. The field operations $+$ and $\odot$ are given by polynomial addition and
polynomial multiplication modulo $p(x)$, respectively. In this work, we exploit $\faktor{\F_q[x]}{p(x)}\cong \F_q^m$, which is obtained by identifying polynomials from the quotient with
their coefficient vectors (row vectors, zero-padded to length $m$ if necessary). It is clear that polynomial
addition in the quotient turns into componentwise addition in $\F_q^m$.

Modular polynomial multiplication
\begin{equation*}
  u(x)\odot v(x)=u(x)v(x) \bmod\,p(x)
\end{equation*}
is slightly more complicated to translate into the vector domain. It is instrumental to separate
polynomial multiplication $w(x)=u(x)v(x)$ and modular reduction $w(x) \bmod\,p(x)$.

Polynomial multiplication coincides with discrete convolution of the coefficient vectors
$\vec{u},\vec{v}\in\F_q^m$, and the latter can be realized by multiplying $\vec{u}$ from the right with a Toeplitz matrix
$\mat{T}_{\vec{v}}\in\F_q^{m\times (2m-1)}$, whose first row consists of the components of $\vec{v}$
followed by $m-1$ zeros. The intermediate result after multiplication is $(w_0, \ldots, w_{m-1}, w_{m},
\ldots, w_{2m-2})=\vec{u}\mat{T}_{\vec{v}}$, which is obviously twice as long as $\vec{u}$ and $\vec{v}$ and therefore \emph{not} an element of $\F_q^m$.
This must be fixed by modular reduction.

Modular reduction with respect to the defining polynomial $p(x)=x^m+\sum_{i=0}^{m-1} p_i x^i$ means
\begin{equation*}
  x^m=-\sum_{i=0}^{m-1} p_ix^i.
\end{equation*}
This allows annihilation of the coefficients $w_{2m-2}, \ldots, w_m$ of $w(x)=\sum_{i=0}^{2m-2}w_ix^i$ one
after the other (starting from the most significant one) by subtracting
\begin{equation*}
  w_j x^j+w_j\sum_{i=0}^{m-1} p_ix^{i+j-m}
\end{equation*}
from $w(x)$ for $j=2m-2, \ldots, m$ (in that particular order). Note that step $j$ can potentially update
the coefficients $w_{j-m}, \ldots, w_{j-1}$ of the respective intermediate result by the rule
\begin{equation*}
  w_{i+j-m}=w_{i+j-m}-w_jp_i,\quad i=0, \ldots, m-1,
\end{equation*}
which can be re-indexed in order to obtain
\begin{equation}\label{eqn:coeffreduction}
  w_\mu=w_\mu-w_jp_{m-j+\mu},\quad \mu=j-m, \ldots, j-1.
\end{equation}

How does this translate into the vector/matrix domain? Each step annihilates the most significant
coefficient of the respective intermediate result, thereby reducing the possible length of its coefficient
vector by one. This means that \eqref{eqn:coeffreduction} is realized by multiplication with a $(j+1)\times
j$ matrix $\mat{R}_j$ over $\F_q$. Coefficients $w_\mu$, $\mu=0, \ldots, j-m-1$, of the intermediate result
remain unaffected, hence columns $\mu=0, \ldots, j-m-1$ of $\mat{R}_j$ must be $\mu$th (column) unit
vectors. Columns $\mu=j-m, \ldots, j-1$ must have one in row $\nu=\mu$ (corresponding to $w_\mu$ in
\eqref{eqn:coeffreduction}), $-p_{m-j+\mu}$ in row $j$ (corresponding to $w_j$ in
\eqref{eqn:coeffreduction}), and zero everywhere else. Consequently, $\mat{R}_j=\begin{bmatrix}R_{\nu,
\mu}\end{bmatrix}$, where
\begin{equation*}
  R_{\nu, \mu}=\left\{\begin{array}{ll}
    1, & \text{if } \nu=\mu\\
    -p_{m-j+\mu}, & \text{if } \nu=j \text{ and } \mu\geq j-m\\
    0, & \text{else}\\
  \end{array}\right..
\end{equation*}

$\mat{R}_j$ consists of an $(j-m)\times (j-m)$ identity matrix $\mat{I}_{j-m}$, a (row) unit vector
$\vec{e}$ of length $m$ and the transposed $m\times m$ companion matrix $\mat{C}^T[p(x)]$ of $p(x)$. That
is,

\begin{equation*}
  \mat{R}_j=
  \begin{tikzpicture}[baseline=0cm]
    \matrix (R) [
      matrix of nodes, nodes in empty cells,
      left delimiter={[},right delimiter={]},
      every node/.style={text width=.71cm, text height=.3cm, font=\footnotesize,
      inner sep=0pt, text depth=.2cm,
      align=center},
      column 9/.style={nodes={text width=1cm}},
      nodes={%
        execute at begin node=$,%
        execute at end node=$%
      }%
    ]
    {
      1 & 0 &   & 0 & 0    &        &   &   & 0\\
      0 &   &   &\\
        &   &   & 0\\
      0 &   & 0  & 1 & 0    &        &   &   & 0\\
      0 &   &   & 0 & 1    & 0      &   &   & 0\\
        &   &   &   & 0    & 1      & 0 &   & 0\\
        &   &   &   &      &        &   &   &\\
        &   &   &   &      &        &   &   & 0\\
        &   &   &   & 0    &        &   & 0 & 1\\
      0 &   &   & 0 & -p_0 & -p_1   &   &   & -p_{m-1}\\
    };

    \draw[loosely dotted] (R-2-1)-- (R-4-1);
    \draw[loosely dotted] (R-5-1)-- (R-10-1);
    \draw[loosely dotted] (R-1-4)-- (R-3-4);
    \draw[loosely dotted] (R-5-4)-- (R-10-4);
    \draw[loosely dotted] (R-1-5)-- (R-4-5);
    \draw[loosely dotted] (R-6-5)-- (R-9-5);
    \draw[loosely dotted] (R-1-9)-- (R-4-9);
    \draw[loosely dotted] (R-6-9)-- (R-8-9);

    \draw[loosely dotted] (R-1-2)-- (R-3-4);
    \draw[loosely dotted] (R-1-1)-- (R-4-4);
    \draw[loosely dotted] (R-2-1)-- (R-4-3);
    \draw[loosely dotted] (R-6-7)-- (R-8-9);
    \draw[loosely dotted] (R-6-6)-- (R-9-9);
    \draw[loosely dotted] (R-6-5)-- (R-9-8);

    \draw[loosely dotted] (R-1-2)-- (R-1-4);
    \draw[loosely dotted] (R-1-5)-- (R-1-9);
    \draw[loosely dotted] (R-4-5)-- (R-4-9);
    \draw[loosely dotted] (R-4-1)-- (R-4-3);
    \draw[loosely dotted] (R-5-1)-- (R-5-4);
    \draw[loosely dotted] (R-5-6)-- (R-5-9);
    \draw[loosely dotted] (R-6-7)-- (R-6-9);
    \draw[loosely dotted] (R-9-5)-- (R-9-8);
    \draw[loosely dotted] (R-10-1)-- (R-10-4);
    \draw[loosely dotted] (R-10-6)-- (R-10-9);

    \draw[red] (R-1-1.north west) -- (R-1-4.north east) -- ([shift={(0, .05)}] R-4-4.south east) --
    ([shift={(0, .05)}] R-4-1.south west) -- cycle;

    \draw[ForestGreen] ([shift={(0, -.05)}] R-5-5.north west) -- ([shift={(0, -.05)}] R-5-9.north east) --
    ([shift={(0, .05)}] R-5-9.south east) -- ([shift={(0, .05)}] R-5-5.south west) -- cycle;

    \draw[blue] ([shift={(0, -.05)}] R-6-5.north west) -- ([shift={(0, -.05)}] R-6-9.north east) -- (R-10-9.south east) --
    (R-10-5.south west) -- cycle;

    \draw[red, decorate, decoration={brace,raise=-2pt, amplitude=5pt}]
      ([yshift=1ex]R-1-1.north west) to node[above] {$\mat{I}_{j-m}$} ([yshift=1ex]R-1-4.north east);

    \draw[ForestGreen, decorate, decoration={brace,raise=-2pt, amplitude=5pt}]
      ([yshift=.3ex]R-5-5.north west) to node[above=1ex] {$\vec{e}$} ([yshift=.3ex]R-5-9.north east);

    \draw[blue, decorate, decoration={brace,mirror,raise=-2pt, amplitude=5pt}]
      ([yshift=-1ex]R-10-5.south west) to node[below] {$\mat{C}^T[p(x)]$} ([yshift=-1ex]R-10-9.south east);
  \end{tikzpicture}.
\end{equation*}

Steps $j=2m-2, \ldots, m$ of modular reduction of the intermediate result $\vec{u}\mat{T}_{\vec{v}}$ can be
performed by multiplication (from the right) with
\begin{equation*}
  \mat{R}=\prod_{j=2m-2}^{m}\mat{R}_j\in\F_q^{(2^m-1)\times m}.
\end{equation*}
Note that $\mat{R}$ is independent of the operands $\vec{u}$ and $\vec{v}$ and can thus be precomputed.

Altogether, modular polynomial multiplication $\odot$ translates to
\begin{equation*}
  \otimes:\left\{\begin{array}{ccl}
    \F_q^m\times \F_q^m & \rightarrow & \F_q^m\\
    \vec{u}, \vec{v} & \mapsto & \vec{u}\mat{T}_{\vec{v}}\mat{R}
  \end{array}\right..
\end{equation*}

With the usual vector addition $+$ over $\F_q$ as additive field operation we have
\begin{equation*}
  \left(\faktor{\F_q[x]}{p(x)}, +, \odot\right)\cong \left(\F_q^m, +, \otimes\right).
\end{equation*}

An element of $\F_Q$ is an element of the subfield $\F_q\subset\F_Q$ if and only if it is constant
(polynomial domain) or if all its components except for the leftmost (least significant) one are zero (vector/matrix
domain).

\section{Message Constraints for Arbitrary GRS Codes}\label{sec:restrict}

Recall that encoding of GRS codes can be accomplished by polynomial evaluation as in \eqref{eqn:code}. It is
well known that the latter can be realized by multiplying the coefficient vector $\vec{f}$ of the message
polynomial $f(x)\in\F_Q^{<k}$ (zero-padded to length $k$ if necessary) with a canonical generator matrix
given by
\begin{multline}\label{eqn:canon}
  \mat{G}=\begin{bmatrix}G_{i, j}\end{bmatrix}=\\%
  \begin{bmatrix}
    b_0 & b_1 & \cdots & b_{n-1}\\
    b_0a_0 & b_1a_1 & \cdots & b_{n-1}a_{n-1}\\
    \vdots & \vdots & \ddots & \vdots\\
    b_0a_0^{k-1} & b_1a_1^{k-1} & \cdots & b_{n-1}a_{n-1}^{k-1}
  \end{bmatrix}\in\F_Q^{k\times n},
\end{multline}
where $\mathcal{A}=(a_0, \ldots, a_{n-1})$ are the locators and $\mathcal{B}=(b_0, \ldots, b_{n-1})$ the
column multipliers of the code $\mathcal{C}$. Recall that $n$ is the length, $k$ the dimension, and
$d=n-k+1$ the minimum distance of $\mathcal{C}$.

\begin{proposition}\label{prop:leading_trailing}
  Let $s, t\in\N$ with $s+t\leq k$. If the coefficients $f_i$, $i\in\{0, \ldots, s-1\}\cup\{k-1-t, \ldots,
  k-1\}$ of \emph{every} message polynomial $f(x)$ are known to be zero then the resulting codewords constitute an
  auxiliary GRS code $\mathcal{C}^\updownarrows$ with length $n^\updownarrows=n$, dimension
  $k^\updownarrows=k-s-t$, minimum distance $d^\updownarrows=d+s+t$, locators
  $\mathcal{A}^\updownarrows=\mathcal{A}$, and column multipliers $\mathcal{B}^\updownarrows=(b_0a_0^s,
  \ldots, b_{n-1}a_{n-1}^s)$.
\end{proposition}

The part about the most significant coefficients is trivial to see: if the coefficients $k-1-t, \ldots,
k-1$ of \emph{every} message polynomial are known to be zero then the last $t$ rows of $\mat{G}$ can
be ignored and the resulting GRS code has length $n$, dimension $k-t$, and minimum distance $d+t$.

If additionally the $s$ least significant coefficients of every message polynomial are zero, then the first
$s$ rows of $\mat{G}$ are superfluous. The resulting codewords can be considered as codewords from an
auxiliary GRS code of length $n$, dimension $k-s-t$ and minimum distance is $d+s+t$. The auxiliary GRS code
$\mathcal{C}^\updownarrows$ has locators $\mathcal{A}$ and column multipliers $(b_0a_0^s, \ldots,
b_{n-1}a_{n-1}^s)$, i.e., its $(k-s-t)\times n$ canonical generator matrix $\mat{G}^\updownarrows$ is
exactly $\mat{G}$ with its first $s$ and last $t$ rows deleted.

We now ask the following question: which constraint on a message $\vec{f}\in\F_Q^k$ has to hold such that
encoding leads to a codeword $\vec{c}=\vec{f}\mat{G}$ from the ``small'' vector space $\F_q^n$ instead of the
``big'' space $\F_Q^n$? Answering this question will provide us with a precise characterization of the
subfield subcode $\mathcal{C}'=\mathcal{C}\cap\F_q^n$, i.e., a generalization of the conjugacy constraints from
\eqref{eqn:conj} for arbitrary GRS codes (and their potentially non-BCH subfield subcodes).

Let us consider encoding with field operations in the vector/matrix domain as elaborated in
Section~\ref{sec:translate}. Vector-matrix multiplication $(c_0, \ldots, c_{n-1})=\vec{f}\mat{G}$ means
calculating
\begin{equation*}
  c_j =\sum_{i=0}^{k-1} f_i G_{i, j},\quad j=0, \ldots, n-1,
\end{equation*}
which, over $\F_q^m$, becomes
\begin{equation*}
  \sum_{i=0}^{k-1} \widetilde{f}_i\otimes G_{i, j}=\sum_{i=0}^{k-1} \widetilde{f}_i
  \mat{T}_{G_{i, j}}\mat{R},\quad j=0, \ldots, n-1.
\end{equation*}

Consequently, if $\vec{f}\in\F_Q^k$ is interpreted as $\widetilde{\vec{f}}\in\F_q^{mk}$, then the generator
matrix becomes
\begin{multline*}
  \widetilde{\mat{G}}=\begin{bmatrix}\mat{T}_{G_{i, j}}\mat{R}\end{bmatrix}\\
  =\begin{bmatrix}
    \mat{T}_{b_0}\mat{R} & \mat{T}_{b_1}\mat{R} & \cdots & \mat{T}_{b_{n-1}}\mat{R}\\
    \mat{T}_{b_0a_0}\mat{R} & \mat{T}_{b_1a_1}\mat{R} & \cdots & \mat{T}_{b_{n-1}a_{n-1}}\mat{R}\\
    \vdots & \vdots & \ddots & \vdots\\
    \mat{T}_{b_0a_0^{k-1}}\mat{R} & \mat{T}_{b_1a_1^{k-1}}\mat{R} & \cdots & \mat{T}_{b_{n-1}a_{n-1}^{k-1}}\mat{R}
  \end{bmatrix}\\
  \in\F_Q^{mk\times mn},
\end{multline*}
i.e., a matrix of matrices.

Now when is the codeword $\widetilde{\vec{c}}=(\widetilde{c}_0, \ldots,
\widetilde{c}_{n-1})=\widetilde{\vec{f}}\widetilde{\mat{G}}$ from $\F_q^n$? As stated at the end of
Section~\ref{sec:translate} this is the case if and only if all the components except for the first one of
all $\widetilde{c}_j$ (when interpreted as a vectors from $\F_q^m$), $j=0, \ldots, n-1$, are zero. This can
be enforced by restricting messages $\widetilde{\vec{f}}$ to the span of a certain matrix
$\widetilde{\mat{\Gamma}}$.

Let $h[\cdot]$ be the function that discards the first column of a matrix. Then a basis matrix
$\widetilde{\mat{\Gamma}}$ of the vector space of all $\widetilde{\vec{f}}$ that are encoded into a
codewords $\widetilde{\vec{c}}\in\F_q^n$ is given by a basis matrix of the kernel of a submatrix of
$\widetilde{\mat{G}}$. It can be obtained by solving the homogeneous linear system
\begin{equation}\label{eqn:system}
  \setlength{\arraycolsep}{2.5pt}
  \left[\begin{array}{cccc|c}
    h[\mat{T}_{b_0}\mat{R}] & h[\mat{T}_{b_0a_0}\mat{R}] & \hdots & h[\mat{T}_{b_0a_0^{k-1}}\mat{R}] & 0\\
    h[\mat{T}_{b_1}\mat{R}] & h[\mat{T}_{b_1a_1}\mat{R}] & \hdots & h[\mat{T}_{b_1a_1^{k-1}}\mat{R}] & 0\\
    \vdots                  & \vdots                     & \ddots & \vdots & \vdots\\
    h[\mat{T}_{b_{n-1}}\mat{R}] & h[\mat{T}_{b_{n-1}a_{n-1}}\mat{R}] & \hdots &
    h[\mat{T}_{b_{n-1}a_{n-1}^{k-1}}\mat{R}] & 0\\
  \end{array}\right]
\end{equation}
over $\F_q$ with $(m-1)n$ equations in $mk$ unknowns. The existence of a non-trivial solution is not guaranteed and
depends on the actual choice of locators $\mathcal{A}$ and column multipliers $\mathcal{B}$. This will be
elaborated in the upcoming example.

Before we start with the example let us provide the answer to our question:
\begin{equation*}
  \vec{c}=\vec{f}\mat{G}\in\F_q^n \Longleftrightarrow \vec{f}\in \operatorname{span}\left[\mat{\Gamma}\right],
\end{equation*}
where $\mat{\Gamma}\in\F_Q^{k'\times k}$ is simply $\widetilde{\mat{\Gamma}}\in\F_q^{k'\times mk}$
interpreted as a $k'\times k$ matrix over $\F_Q$. We have the following theorem:

\begin{theorem}\label{thm:ssc}
  If $\mathcal{C}$ is a GRS code over $\F_{q^m}$ with length $n$, dimension $k$, minimum distance $d$,
  locators $\mathcal{A}=(a_0, \ldots, a_{n-1})$, and column multipliers $\mathcal{B}=(b_0, \ldots,
  b_{n-1})$, then its subfield subcode $\mathcal{C}'=\mathcal{C}\cap\F_q^n$ has generator matrix
  $\mat{G}'=\mat{\Gamma}\mat{G}\in\F_q^{k'\times n}$, where $\mat{\Gamma}$ is obtained as basis matrix of the
  solution space of \eqref{eqn:system}.  The dimension of $\mathcal{C}'$ is
  $k'=\operatorname{rank}[\mat{\Gamma}]$ and the design distance is $d'=d^\updownarrows$, where
  $d^\updownarrows$ is obtained using Proposition~\ref{prop:leading_trailing}.

\end{theorem}

Note that $d^\updownarrows$ depends on $s$, $t$ (and thereby also on
$\mat{\Gamma}$), which is not reflected in the $\updownarrows$ notation.

\begin{example}
Let $\mathcal{C}$ be the cyclic GRS code of length $n=7$, dimension $k=5$ and minimum distance $d=3$ over
$\F_{2^3}$ (defining polynomial $p(x)=x^3+x+1$, $m=3$) with parameter $\delta=0$ (cf.
Section~\ref{sec:codes}). The corresponding generator matrix $\widetilde{\mat{G}}$ is

\begin{equation}\label{eqn:Gtilde}
  \widetilde{\mat{G}}{=}
  \begin{tikzpicture}[baseline=0ex]
    \matrix (G) [
      matrix of nodes, nodes in empty cells,
      left delimiter={[},right delimiter={]},
      every node/.style={font=\footnotesize}, inner sep=2.9pt,
      nodes={%
        execute at begin node=$,%
        execute at end node=$%
      }%
    ]
    {
    1 & 0 & 0 & 1 & 0 & 0 & 1 & 0 & 0 & 1 & 0 & 0 & 1 & 0 & 0 & 1 & 0 & 0 & 1 & 0 & 0\\
    0 & 1 & 0 & 0 & 1 & 0 & 0 & 1 & 0 & 0 & 1 & 0 & 0 & 1 & 0 & 0 & 1 & 0 & 0 & 1 & 0\\
    0 & 0 & 1 & 0 & 0 & 1 & 0 & 0 & 1 & 0 & 0 & 1 & 0 & 0 & 1 & 0 & 0 & 1 & 0 & 0 & 1\\
    1 & 0 & 0 & 0 & 1 & 0 & 0 & 0 & 1 & 1 & 1 & 0 & 0 & 1 & 1 & 1 & 1 & 1 & 1 & 0 & 1\\
    0 & 1 & 0 & 0 & 0 & 1 & 1 & 1 & 0 & 0 & 1 & 1 & 1 & 1 & 1 & 1 & 0 & 1 & 1 & 0 & 0\\
    0 & 0 & 1 & 1 & 1 & 0 & 0 & 1 & 1 & 1 & 1 & 1 & 1 & 0 & 1 & 1 & 0 & 0 & 0 & 1 & 0\\
    1 & 0 & 0 & 0 & 0 & 1 & 0 & 1 & 1 & 1 & 0 & 1 & 0 & 1 & 0 & 1 & 1 & 0 & 1 & 1 & 1\\
    0 & 1 & 0 & 1 & 1 & 0 & 1 & 1 & 1 & 1 & 0 & 0 & 0 & 0 & 1 & 0 & 1 & 1 & 1 & 0 & 1\\
    0 & 0 & 1 & 0 & 1 & 1 & 1 & 0 & 1 & 0 & 1 & 0 & 1 & 1 & 0 & 1 & 1 & 1 & 1 & 0 & 0\\
    1 & 0 & 0 & 1 & 1 & 0 & 1 & 0 & 1 & 0 & 0 & 1 & 1 & 1 & 1 & 0 & 1 & 0 & 0 & 1 & 1\\
    0 & 1 & 0 & 0 & 1 & 1 & 1 & 0 & 0 & 1 & 1 & 0 & 1 & 0 & 1 & 0 & 0 & 1 & 1 & 1 & 1\\
    0 & 0 & 1 & 1 & 1 & 1 & 0 & 1 & 0 & 0 & 1 & 1 & 1 & 0 & 0 & 1 & 1 & 0 & 1 & 0 & 1\\
    1 & 0 & 0 & 0 & 1 & 1 & 0 & 1 & 0 & 1 & 1 & 1 & 0 & 0 & 1 & 1 & 0 & 1 & 1 & 1 & 0\\
    0 & 1 & 0 & 1 & 1 & 1 & 0 & 0 & 1 & 1 & 0 & 1 & 1 & 1 & 0 & 1 & 0 & 0 & 0 & 1 & 1\\
    0 & 0 & 1 & 1 & 0 & 1 & 1 & 1 & 0 & 1 & 0 & 0 & 0 & 1 & 1 & 0 & 1 & 0 & 1 & 1 & 1\\
    };

     \def\bshrink{.2}
     \def\rshrink{.4}

    \foreach \x in {0, ..., 6} {
      \pgfmathtruncatemacro{\tempa}{3*\x+1}
      \pgfmathtruncatemacro{\tempb}{3*\x+3}
      \pgfmathtruncatemacro{\tempe}{3*\x+2}
      \foreach \y in {0, ..., 4} {
        \pgfmathtruncatemacro{\tempc}{3*\y+1}
        \pgfmathtruncatemacro{\tempd}{3*\y+3}

        \draw[dotted, blue]
          ([xshift=\bshrink ex, yshift=-\bshrink ex] G-\tempc-\tempa.north west) --
          ([xshift=-\bshrink ex, yshift=-\bshrink ex] G-\tempc-\tempb.north east) --
          ([xshift=-\bshrink ex, yshift=\bshrink ex] G-\tempd-\tempb.south east) --
          ([xshift=\bshrink ex, yshift=\bshrink ex] G-\tempd-\tempa.south west) --
          cycle;

          \draw[red]
          ([xshift=\rshrink ex, yshift=-\rshrink ex] G-\tempc-\tempe.north west) --
          ([xshift=-\rshrink ex, yshift=-\rshrink ex] G-\tempc-\tempb.north east) --
          ([xshift=-\rshrink ex, yshift=\rshrink ex] G-\tempd-\tempb.south east) --
          ([xshift=\rshrink ex, yshift=\rshrink ex] G-\tempd-\tempe.south west) --
          cycle;
      }
    }

    \draw[red, decorate, decoration={brace, raise=-2pt, amplitude=5pt}]
      ([xshift=\rshrink ex, yshift=.5ex] G-1-2.north west) to node[above] {\small$h[\mat{T}_{b_0}\mat{R}]$} ([xshift=-\rshrink ex, yshift=.5ex]G-1-3.north east);

    \draw[blue, decorate, decoration={brace, raise=-2pt, amplitude=5pt}]
      ([xshift=\bshrink ex, yshift=4ex] G-1-1.north west) to node[above] {\small$\mat{T}_{b_0}\mat{R}$} ([xshift=-\bshrink ex, yshift=4ex]G-1-3.north east);

  \end{tikzpicture},
\end{equation}

where the $\mat{T}_{G_{i, j}}\mat{R}$ blocks are marked by dotted blue boxes and the $h[\mat{T}_{G_{i, j}}\mat{R}]$
blocks by red boxes for clarity.

Setting up and solving the linear system \eqref{eqn:system} delivers the $4\times (3\cdot 5)$ basis matrix
\begin{equation}\label{eqn:ex1_1}
  \setlength{\arraycolsep}{2.5pt}
  \widetilde{\mat{\Gamma}}=\left[\begin{array}{ccc:ccc:ccc:ccc:ccc}
  1 & 0 & 0 & 0 & 0 & 0 & 0 & 0 & 0 & 0 & 0 & 0 & 0 & 0 & 0\\
  0 & 0 & 0 & 1 & 0 & 0 & 1 & 0 & 0 & 0 & 0 & 0 & 1 & 0 & 0\\
  0 & 0 & 0 & 0 & 1 & 0 & 0 & 0 & 1 & 0 & 0 & 0 & 0 & 1 & 1\\
  0 & 0 & 0 & 0 & 0 & 1 & 0 & 1 & 1 & 0 & 0 & 0 & 0 & 1 & 0\\
  \end{array}\right]
\end{equation}
over $\F_2$. Thus, the dimension of the subfield subcode in the case $\delta=0$ is
$k'=\operatorname{rank}[\widetilde{\mat{\Gamma}}]=4$. There are no zero columns neither on the left nor on
the right of the matrix, meaning that $\mathcal{C}^\updownarrows=\mathcal{C}$ and consequently the
design distance is $d'=d^\updownarrows=d=3$.

Choosing parameter $\delta=1$ instead results in  the $3\times (3\cdot 5)$ basis matrix
\begin{equation}\label{eqn:ex1_2}
  \setlength{\arraycolsep}{2.5pt}
  \widetilde{\mat{\Gamma}}=\left[\begin{array}{ccc:ccc:ccc:ccc:ccc}
  1 & 0 & 0 & 1 & 0 & 0 & 0 & 0 & 0 & 1 & 0 & 0 & 0 & 0 & 0\\
  0 & 1 & 0 & 0 & 0 & 1 & 0 & 0 & 0 & 0 & 1 & 1 & 0 & 0 & 0\\
  0 & 0 & 1 & 0 & 1 & 1 & 0 & 0 & 0 & 0 & 1 & 0 & 0 & 0 & 0\\
  \end{array}\right],
\end{equation}
and in that case $k'=3$. The $m=3$ rightmost columns of this matrix are zero, which, when translated into
polynomial domain, means that the $t=\lfloor\nicefrac{3}{m}\rfloor=1$ most significant coefficients of
every message polynomial fulfilling the constraint are zero. There are no zero columns on the left ($s$ is
zero). We can apply Proposition~\ref{prop:leading_trailing} in order to obtain design distance
$d'=d^\updownarrows=4\geq d$.

Choosing $\delta=4$ gives the $1\times (3\cdot 5)$ basis matrix
\begin{equation}\label{eqn:ex1_3}
  \setlength{\arraycolsep}{2.5pt}
  \widetilde{\mat{\Gamma}}=\left[\begin{array}{ccc:ccc:ccc:ccc:ccc}
  0 & 0 & 0 & 0 & 0 & 0 & 0 & 0 & 0 & 1 & 0 & 0 & 0 & 0 & 0\\
  \end{array}\right],
\end{equation}
and we have $k'=1$. The $9$ leftmost columns and the $5$ rightmost columns are zero. Thus, we have
$s=\lfloor\nicefrac{9}{m}\rfloor=3$ and $t=\lfloor\nicefrac{5}{m}\rfloor=1$ and, again using
Proposition~\ref{prop:leading_trailing}, design distance $d'=d^\updownarrows=7\geq d$.
\end{example}

Note that the fact that $\mathcal{C}$ is cyclic is not required for obtaining the subfield subcodes. A
cyclic code was chosen for the example because tables of BCH codes are widely available for comparison,
e.g. in \cite{macwilliams_sloane:1988}.

The example suggests using Proposition~\ref{prop:leading_trailing} in order to obtain subcodes of subfield
subcodes with increasing design distance. This is subject of the following section.

\section{Nested Subfield Subcodes}\label{sec:families}

Finding nested subcodes with increasing distance is conceptually simple. Nevertheless, a full
algorithmic description is very technical. Due to space restrictions we can only give a coarse overview of the
procedure in this section and refer to the upcoming full paper.

Consider the basis matrix $\widetilde{\mat{\Gamma}}\in\F_q^{k'\times mk}$ and, w.l.o.g., assume it is in
reduced row echelon form. We are free to remove $u$ rows from $\widetilde{\mat{\Gamma}}$, leading (cf. Theorem~\ref{thm:ssc}) to dimension $k'-u$.

How does this affect the design distance? In the general case not at all. Consider for example
$\widetilde{\mat{\Gamma}}$ from \eqref{eqn:ex1_2} and remove its first row. This reduces the dimension of
the subfield subcode to $k'-1=2$, but the design distance stays at $d'=d^\updownarrows=4$. Removing the
row only makes the code worse.

If on the other hand we remove the first row of $\widetilde{\mat{\Gamma}}$ from \eqref{eqn:ex1_1} we obtain
\begin{equation*}
  \setlength{\arraycolsep}{2.5pt}
  \widetilde{\mat{\Gamma}}=\left[\begin{array}{ccc:ccc:ccc:ccc:ccc}
  0 & 0 & 0 & 1 & 0 & 0 & 1 & 0 & 0 & 0 & 0 & 0 & 1 & 0 & 0\\
  0 & 0 & 0 & 0 & 1 & 0 & 0 & 0 & 1 & 0 & 0 & 0 & 0 & 1 & 1\\
  0 & 0 & 0 & 0 & 0 & 1 & 0 & 1 & 1 & 0 & 0 & 0 & 0 & 1 & 0\\
  \end{array}\right].
\end{equation*}
Without doubt, the dimension becomes $k'-1=3$. But what about the design distance? Note the block of $3$
zero columns on the left! They allow us to invoke Proposition~\ref{prop:leading_trailing} in order to get an
auxiliary code $\mathcal{C}^\updownarrows$ with $d^\updownarrows=d+\lfloor\nicefrac{3}{m}\rfloor=4$,
increasing the design distance to $d'=4$. The striking property of this code is, that it is a true subcode
of $\mathcal{C}'$ \emph{and} that its design distance is larger than that of $\mathcal{C}'$.

In order to find \emph{all} nested subfield subcodes for given locators and column multipliers, we have to start with the largest
possible GRS code, namely the trivial one with dimension $k=n$ and minimum distance $d=1$.

\begin{example}
Let $\mathcal{C}$ be the cyclic GRS code of length $n=7$, dimension $k=7$ and minimum distance $d=1$ over
$\F_{2^3}$ (defining polynomial $p(x)=x^3+x+1$, $m=3$) with parameter $\delta=0$ (cf.
Section~\ref{sec:codes}). 

Setting up and solving the linear system \eqref{eqn:system} delivers the $7\times (3\cdot 7)$ basis matrix
shown in \eqref{eqn:Gammatilde}. The resulting subfield subcode has dimension $k'=7$ and design distance
$d'=d=1$. With reference to Proposition~\ref{prop:leading_trailing}, leading and trailing groups of $m=3$
zeros in each row are separated from the center of the matrix by an $s$ and $t$ trajectory, respectively.

Removing the last three rows results in the two nested codes from the beginning of this section (dimension
$4$, design distance $3$ and dimension $3$, design distance $4$, respectively. Removing the first four
rows results in a subfield subcode with $k'=3$ and design distance $d'=4$.

\begin{figure*}[!b] % !b requires \usepackage{stfloats}
% The spacer can be tweaked to stop underfull vboxes.
%\vspace*{4pt}
% IEEE uses as a separator
\hrulefill
% ensure that we have normalsize text
\normalsize
\begin{equation}\label{eqn:Gammatilde}
  \widetilde{\mat{\Gamma}}=
  \begin{tikzpicture}[baseline=0ex]
    \matrix (G) [
      matrix of nodes, nodes in empty cells,
      left delimiter={[},right delimiter={]},
      every node/.style={font=\footnotesize}, inner sep=5pt,
      nodes={%
        execute at begin node=$,%
        execute at end node=$%
      }%
    ]
    {
    1 & 0 & 0 & 0 & 0 & 0 & 0 & 0 & 0 & 0 & 0 & 0 & 0 & 0 & 0 & 0 & 0 & 0 & 0 & 0 & 0\\
    0 & 0 & 0 & 1 & 0 & 0 & 1 & 0 & 0 & 0 & 0 & 0 & 1 & 0 & 0 & 0 & 0 & 0 & 0 & 0 & 0\\
    0 & 0 & 0 & 0 & 1 & 0 & 0 & 0 & 1 & 0 & 0 & 0 & 0 & 1 & 1 & 0 & 0 & 0 & 0 & 0 & 0\\
    0 & 0 & 0 & 0 & 0 & 1 & 0 & 1 & 1 & 0 & 0 & 0 & 0 & 1 & 0 & 0 & 0 & 0 & 0 & 0 & 0\\
    0 & 0 & 0 & 0 & 0 & 0 & 0 & 0 & 0 & 1 & 0 & 0 & 0 & 0 & 0 & 1 & 0 & 0 & 1 & 0 & 0\\
    0 & 0 & 0 & 0 & 0 & 0 & 0 & 0 & 0 & 0 & 1 & 0 & 0 & 0 & 0 & 0 & 1 & 1 & 0 & 0 & 1\\
    0 & 0 & 0 & 0 & 0 & 0 & 0 & 0 & 0 & 0 & 0 & 1 & 0 & 0 & 0 & 0 & 1 & 0 & 0 & 1 & 1\\
    };

    \foreach \x in {0, ..., 5} {
      \pgfmathtruncatemacro{\temp}{3*\x+3}

       \draw[loosely dashed] (G-1-\temp.north east) -- (G-7-\temp.south east);
    }

    \draw[thick, red] ([xshift=-.5ex] G-1-1.north west) -- ([xshift=-.5ex] G-1-1.south west) -- ([xshift=-.5ex] G-1-4.south west) -- ([xshift=-.5ex] G-4-4.south west) --
    ([xshift=-.5ex] G-4-10.south west) -- ([xshift=-.5ex] G-7-10.south west);
    \node[red, anchor=south west] at ([shift={(-1ex, 1.2ex)}] G-1-1.north west) {$s$ trajectory};

    \draw[thick, blue] ([xshift=-.5ex] G-1-16.north west) -- ([xshift=-.5ex] G-4-16.south west) -- ([xshift=-.5ex] G-4-21.south east) -- ([xshift=-.5ex] G-7-21.south east);
    \node[blue, anchor=south west] at ([shift={(-1ex, 1.2ex)}] G-1-16.north west) {$t$ trajectory};

    \draw[decorate, decoration={brace,raise=-2pt, amplitude=5pt}]
      ([shift={(.15, 1ex)}] G-1-1.north west) to node[above] {$\widetilde{\mat{\Gamma}}$ from
      \eqref{eqn:ex1_1}} ([shift={(-.15, 1ex)}] G-1-15.north east);

    \draw ([shift={(.15, -.1)}] G-1-1.north west) -- ([shift={(-.15, -.1)}] G-1-15.north east) --
    ([shift={(-.15, .1)}] G-4-15.south east) -- ([shift={(.15, .1)}] G-4-1.south west) -- cycle;
    
  \end{tikzpicture}.
\end{equation}
% The spacer can be tweaked to stop underfull vboxes.
\vspace*{-7pt}
\end{figure*}

\end{example}

In general, the procedure for finding nested subcodes with increasing design distance for
arbitrary but fixed locators $\mathcal{A}$ and column multipliers $ \mathcal{B}$ can be outlined as follows.
\begin{enumerate}[(1)]
  \item Calculate $\widetilde{\mat{\Gamma}}$ for the GRS code with $k=n$.
  \item Pick any submatrix $\mat{M}$ with $k'$ rows, such that
    \begin{enumerate}[(a)]
       \item its first row is bounded by the $s$ trajectory on the left,
       \item its last row is bounded by the $t$ trajectory on the right, and
       \item $\widetilde{\mat{\Gamma}}$ must not have any nonzero components left of the $s$ and right of the $t$ trajectory in
         the rows from which $\mat{M}$ is taken.
     \end{enumerate}
  \item Starting from the top, remove rows from $\mat{M}$. Every row removed decreases $k'$ of the nested
    subcode by one. The design distance of the current nested code coincides with the number of columns located completely
    outside of the $s$ and $t$ trajectories (counted in groups of $m$).
\end{enumerate}

\section{Conclusion}

Constructing BCH codes (subcodes of cyclic GRS codes) based on minimal polynomials and calculating their
design distance based on consecutive zeros in their generator polynomials is textbook knowledge. We
provided a more general approach, which can deal with \emph{arbitrary} GRS codes and their alternant
subfield subcodes. Our approach requires nothing else than linear algebra over finite fields, which we
believe is an advantage in its own right.

Searching for ``good'' nested subfield subcodes is particularly simple using an algorithm based
on the $s$ and $t$ trajectories from Section~\ref{sec:families}. An upcoming paper will provide tables of
such codes for practically relevant code parameters.

Even though we restricted ourselves to cyclic GRS  codes and canonical generator matrices in the examples
(since the resulting codes are well known), this is not a restriction of the
approach itself. It can also be applied to, e.g., systematic generator matrices and we can hope for finding
nested subfield subcodes with systematic encoders that way. One track of ongoing research is
further generalization wrt. the representation of $\F_Q$, another one is applying the approach to locally
recoverable codes (LRC) codes, which can also be interpreted as GRS codes with message constraints.

\bibliography{ssc.bbl}
%\bibliographystyle{etc/IEEEtran_doi}
%\bibliography{bib/IEEEabrv,bib/chsenger-paper_ssc}

\end{document}